\documentclass[conference]{IEEEtran}
\IEEEoverridecommandlockouts

\usepackage{cite}
\usepackage{amsfonts}
\usepackage{textcomp}
\usepackage{epsfig}
\usepackage{calc}
\usepackage{amssymb}
\usepackage{amstext}
\usepackage{amsmath}
\usepackage{amsthm}
\usepackage{multicol}
\usepackage{pslatex}
\usepackage[]{algorithm2e}
\usepackage{multirow}
\usepackage{graphicx}
\usepackage{color}
\usepackage{subcaption}
\usepackage[mathscr]{euscript}

\def\BibTeX{{\rm B\kern-.05em{\sc i\kern-.025em b}\kern-.08em
    T\kern-.1667em\lower.7ex\hbox{E}\kern-.125emX}}
    
\newtheorem{definition}{Definition}

\newtheorem{proposition}{Proposition}
\newcommand{\cmmnt}[1]{\ignorespaces}
\newcommand\Tstrut{\rule{0pt}{2.6ex}}       
\newcommand\Bstrut{\rule[-0.9ex]{0pt}{0pt}} 
\newcommand{\TBstrut}{\Tstrut\Bstrut} 

\SetCommentSty{mycommfont}

\makeindex

\begin{document}

\title{Maintaining Ad-Hoc Communication Network in Area Protection Scenarios with Adversarial Agents
}

\author{\IEEEauthorblockN{1\textsuperscript{st} Marika Ivanov\'{a}}
\IEEEauthorblockA{\textit{Department of Informatics} \\
\textit{University of Bergen}\\
HIB - Thormøhlensgt. 55, 5020 Bergen, Norway \\
marika.ivanova@uib.no}	
\and
\IEEEauthorblockN{2\textsuperscript{nd} Pavel Surynek~~~~3\textsuperscript{rd} Diep Thi Ngoc Nguyen}
\IEEEauthorblockA{\textit{Artificial Intelligence Research Center} \\
\textit{National Institute of Advanced }\\
\textit{Industrial Science and Technology (AIST) }\\
2-3-26, Aomi, Koto-ku, Tokyo 135-0064, Japan \\
pavel.surynek@aist.go.jp, diep.nguyen@aist.go.jp}
}

\maketitle

\begin{abstract}
We address a problem of area protection in graph-based scenarios with multiple mobile agents where connectivity is maintained among agents to ensure they can communicate. The problem consists of two adversarial teams of agents that move in an undirected graph shared by both teams. Agents are placed in vertices of the graph; at most one agent can occupy a vertex; and they can move into adjacent vertices in a conflict free way. Teams have asymmetric goals: the aim of one team - {\em attackers} - is to invade into given area while the aim of the opponent team - {\em defenders} - is to protect the area from being entered by attackers by occupying selected vertices. The team of defenders need to maintain connectivity of vertices occupied by its own agents in a visibility graph. The visibility graph models possibility of communication between pairs of vertices.

We study strategies for allocating vertices to be occupied by the team of defenders to block attacking agents where connectivity is maintained at the same time. To do this we reserve a subset of defending agents that do not try to block the attackers but instead are placed to support connectivity of the team. The performance of strategies is tested in multiple benchmarks. The success of a strategy is heavily dependent on the type of the instance, and so one of the contributions of this work is that we identify suitable strategies for diverse instance types.
\end{abstract}


\section{\uppercase{Introduction}}
\label{sec:introduction}
\noindent
\begin{figure}[!h]
  \centering
   {\epsfig{file = 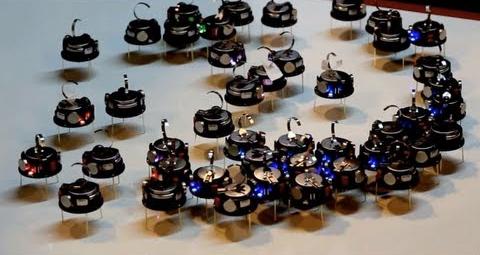, width = 7.0cm}}
  \caption{An example showing Kilobots \cite{RubensteinAN12} - small mobile robots communicating through blinking LEDs and IR sensors}
  \label{fig:kilobot}
 \end{figure}
In this work we study a generalization of {\em Area Protection Problem} (APP) with connectivity maintenance (APPC). In addition to APP, where two teams of mobile agents move in an undirected graph in a conflict free way, connectivity of the set of occupied vertices with respect to a visibility graph is required in APPC. The visibility graph is derived from the graph in which agents move; it has the same set of vertices, but the set of edges is different in general.

APP itself can be regarded as a modification of known problem of {\em Adversarial Cooperative Path Finding} (ACPF) \cite{IvanovaS14} where two teams of agents compete in reaching their target positions. Unlike ACPF, where the goals of teams of agents are symmetric - agents of each team try to reach their targets as first, the adversarial teams in APP have different objectives. The first team of {\em attackers} contains agents whose goal is to reach a pre-defined target location in the area being protected by the second team of {\em defenders}. Each attacker has a unique target in the protected area and each target is assigned to exactly one attacker. The opponent team of defenders tries to prevent the attackers from reaching their targets by occupying selected locations so that they cannot be passed by attackers. Specially in APPC, we require that vertices occupied by the defender team always form a connected subgraph with respect to the visibility graph.

The common feature of APP and APPC is that once a location is occupied by an agent it cannot be entered by another agent until it is first vacated by the agent which occupies it (opposing agent cannot push it out). This property represents a key tool for the defenders to protect the area.

APPC has many real-life motivations from the domains of access denial operations, robotics with adversarial teams of robots or other type of penetrators \cite{DBLP:journals/jair/AgmonKK11}, and computer games. In most practical applications, agents of given team need to communicate with each other while individual robots can communicate at short visual range only as it has been already done in contemporary multi-robot systems (see Figure \ref{fig:kilobot}). Hence it needs to be ensured that the communication reaches every agent of the team. Such property can be modeled as connectivity over the visibility graph whose edges represent possibility of communication between pairs of vertices.

Our contribution consists in suggesting several on-line solving strategies for defenders that allocate suitable vertices to be occupied so that attacker agents cannot pass into the protected area and connectivity in the defender team is maintained. We identified suitable vertex allocation strategies for diverse types of APPC instances and tested them experimentally.

\subsection{Related Work}
APPC, APP, as well as ACPF share the way how movement of agents is treated with the basic variant of {\em cooperative path-finding problem - CPF} ({\em multi-agent path-finding} - MAPF) \cite{Silver05,Ryan08,WangB11}. In CPF the task is to plan movement of agents so that each agent reaches its unique target in a conflict free manner. Movements of agents in APPC at the low reactive level are assumed to be planned by some CPF algorithm where agents of own team cooperate while opposing agents are considered as obstacles.

There exist multiple CPF algorithms both complete and incomplete as well as optimal and sub-optimal under various objective functions. A good compromise between quality of solutions and the speed of solving is represented by suboptimal/incomplete search based methods which are derived from the standard {\em A*} algorithm. These methods include {\em LRA*}, {\em CA*}, {\em HCA*}, and {\em WHCA*} \cite{Silver05}. They provide solutions where individual paths of agents tend to be close to respective shortest paths connecting agents' locations and their targets. Conflict avoidance among agents is implemented via a so called reservation table in case of {\em CA*}, {\em HCA*}, and {\em WHCA*} while {\em LRA*} relies on replanning whenever a conflict occurs. Since our setting in APPC is inherently suitable for a replanning algorithm, {\em LRA*} is a candidate for underlying CPF algorithm for APPC. Moreover {\em LRA*} is scalable for large number of agents which is expected to happen in APPC.

Aside from CPF algorithms, systems with mobile agents that act in the adversarial manner represent another related area. These studies often focus on patroling strategies that are robust with respect to various attackers trying to penetrate through the patrol path \cite{DBLP:journals/amai/ElmaliachAK09}. Theoretical works related to APP also include studies on {\em pursuit evasion} \cite{DBLP:journals/trob/VidalSKSS02} or {\em predator-prey} \cite{DBLP:conf/ijcai/HaynesS95} problems. The major difference between these works and the concept of APP/APPC is that we consider relatively higher number of agents and our agents are more limited in their abilities.

\subsection{Task Decomposition in APPC}

As APPC represents generalization of APP which itself is computationally hard problem \cite{IvanovaSurynek17} (namely PSPACE-hard), we suggest to decompose APPC from the defender's perspective into two different sub-problems: {\em target allocation} and {\em communication maintenance}.

\subsubsection{Target Allocation Problem}

This is the major subproblem being addressed in APP. The defenders are initially not assigned to any targets and don't have any information about the intended target of any attacker. However, the defenders have a full knowledge of all target locations in the protected area. The task in this setting is to allocate each defender agent to some location in the graph so that via its occupation defenders try to optimize a given objective function.

We assume that both teams use the same {\em cooperative path-finding} (CPF) algorithm for reaching temporarily selected targets. Generally, targets can be reassigned multiple times to defender agents in the course of area protection. However, it is assumed that target reassignment does not occur often. After assigning defender agents their target locations they will proceed to their targets via given CPF algorithm. If a target location is reached by a defender agent the agent stops there and continue in occupation of the target location until a new target is assigned to the agent. Attacker agents have their fixed targets in the protected area however they are free to select any temporary target which allows them to move freely in principle.

\subsubsection{Communication Maintenance}

APPC requires, in addition to APP, that any two defenders are able to communicate with each other at any time during their movement. In various practical applications of APP, the possibility of sending messages among the agents is often demanded. Agents may be equipped by an omnidirectional antenna or visual communication device (such as LEDs and IR sensors \cite{RubensteinAN12}), and hence a message reaches all nodes within the communication range of its sender. This feature is often referred to as wireless advantage \cite{Wieselthier}. We assume that the agents have equal and constant communication range, and that they can also work as transceivers, which means that they have the ability to both transmit and receive a signal.

\subsection{Contribution}
Our effort is to design target allocation strategies for the defending team. A hard constraint that can never be violated will be that defending agents always form a connected component. Success of the strategy will measured from the defenders' perspective via an objective function which plays a role of soft constraint (area protection may not be perfect). The following objective functions can be pursued:

\begin{enumerate}
\item maximize the number of target locations that are not captured by the corresponding attacker
\item maximize the number of target locations that are not captured by the corresponding attacker within a given time limit
\item maximize the sum of distances between the attackers and their corresponding targets
\item minimize the time spent at captured targets
\end{enumerate}

\section{\uppercase{definitions and assumptions}}

In APP, we model the environment by an undirected unweighted graph $G=(V,E)$. In this work we restrict the instances to 4-connected grid graphs with possible obstacles. The team of attackers and defenders is denoted by $A=\{a_1,\dots, a_m\}$ and $D=\{d_1,\dots d_n\}$, respectively. 
Continuous time is divided into discrete time steps. At each time step agents are placed in vertices of the graph so that at most one agent is placed in each vertex. Let $\alpha_t: A \cup D\rightarrow V$ be a uniquely invertible mapping denoting configuration of agents at time step $t$.
Agents can wait or move instantaneously into adjacent vertex between successive time steps to form the next configuration $\alpha_{t+1}$. Abiding by the following movement rules ensures preventing conflicts:
\begin{itemize}
\item An agent can move to an adjacent vertex only if the vertex is empty, or is being left at the same time step by another agent
\item No two agents enter the same adjacent vertex at the same time
\item A pair of agents cannot swap across an edge
\end{itemize}

We do not assume any specific order in which agents perform their conflict free actions at each time step. However, our experimental implementation moves all attacking agents prior to moving all defender agents at each time step.
The mapping $\delta^A: A\rightarrow V$ assigns a unique target to each attacker. The task in APP is to move defender agents so that area specified by $\delta^A$ is protected. This task can be equivalently specified as a search for strategy of target assignments for the defender team. That is, we are trying to find an injective mapping $\delta_{t}^D : D\rightarrow V$ which specifies where defender agents should proceed via given path-finding algorithm at time step $t$ as a response to previous attackers movements. The superscripts $A$ and $D$ is sometimes dropped when there is no danger of confusion. Let us note that target reassignment can be done at each time step which is equivalent to full control of movements of defender agents at each time step.

Formally, we state the APP as a decision problem and an optimization problem as follows:

\begin{definition} The decision APP problem:
Given an instance $\Sigma = (G, A, D, \alpha_0, \delta^A)$ of APP, is there a strategy of target allocations $\delta_{t}^D : D\rightarrow V$ such that the team $D$ of defenders is able to prevent agents from the team of attackers from reaching their targets by moving defending agents towards $\delta_{t}^D$.
\end{definition}

In many instances it is not possible to protect all targets. We are therefore also interested in the optimization variant of the APP problem:

\begin{definition} The optimization problem
Given an instance $\Sigma = (G, A, D, \alpha_0, \delta^A)$ of APP, the task is to find a strategy of target allocations $\delta_{t}^D : D\rightarrow V$ such that the team $D$ of defenders minimizes the number of attackers that reach their target by moving defending agents towards $\delta_{t}^D$.
\end{definition}

\subsection{From APP to APPC}
APPC generalizes APP by considering connectivity constraint. As we assume that $G$ is always a grid graph we can introduce connectivity constraint in the following way.

Consider an embedding of $G$ in a plane such that all edges have length 1 and each vertex $v\in V$ has coordinates $(x_v, y_v)$. The physical location $l_v$ represented by $v$ is the unit square area centered at $C_v=(x_v, y_v)$. Furthermore, let $O$ denote the set of square locations representing obstacles.  

Let $r$ be the visibility range, i.e. the maximum distance between two locations such that two agents located at them can communicate together. The locations $l_u$ and $l_v$ can communicate with each other if the line segment $C_uC_v$ does not intersect any obstacle and the length of the shortest path $p_{uv}$ from $u$ to $v$ is at most $r$; shortly we say that $l_u$ is visible from $l_v$ and vice-versa. The visibility graph $G_r=(V,E_r)$ for a visibility range $r$ contains edges between every two vertices that are mutually visible, formally: $(u,v)\in E_r\Leftrightarrow C_uC_v\cap O=\emptyset\wedge |p_{uv}|\leq r$. For any $S\subseteq V$ we use $G_r\left[S\right]$ in order to denote a subgraph of $G_r$ induced by $S$.

\section{\uppercase{Target allocation}}

Since solving APP in practice is a challenging problem due to its high computational complexity, designed methods are inexact and heuristic. Solving approaches can be divided into two basic categories: \emph{single-stage} and \emph{multi-stage}. In single stage methods, targets are assigned to defenders only once at the beginning, as opposed to multi-stage methods, where the targets can be reassigned any time during the agents' course. Once all defenders are allocated to some targets, they try to get to the desired locations using the LRA* algorithm modified for the environment with adversarial team. This work focuses merely on the single-stage methods. In all the studied strategies, every agent is allocated to exactly one location and every location is assigned to at most one defender. We describe several simple target allocation strategies and discuss their properties. 

\subsection{Random Allocation}

For the sake of comparison, we consider a strategy, where each defender is allocated to a random target of an attacker. Neither the agent location nor the underlying grid graph structure is exploited.

\subsection{Greedy Allocation}

A greedy strategy is slightly improved approach. It takes the defenders one by one in a random order and allocates them to their closest target. Greedy as well as Random strategy do not consider initial locations of attackers and do not exploit the structure of underlying graph in any way. These two methods always allocate defenders to given targets of attackers. The advantage of this approach is that if a defender manages to reach its assigned target, it will never be captured by the attacker aiming for that target. This can be useful in scenarios where the number of defenders is similar to the number of attackers. Unfortunately, such a  strategy would not be very successful in instances where attackers significantly outnumber defenders.

\subsection{Bottleneck Simulation Allocation}

\begin{figure}[!h]
  \centering
   {\epsfig{file = 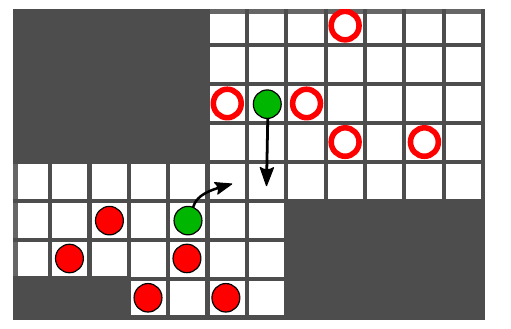, width = 4.0cm}}
  \caption{An example of bottleneck blocking. The defenders (green circles) may protect all the targets (empty red circles) from attackers (red circles) if they move to locations marked by the two arrows. }
  \label{fig:bottleneck}
 \end{figure}

The idea behing the bottleneck simulation strategy is to gain some information from the map structure and the positions of attackers and assign defenders to vertices that would divert attackers from the protected area as much as possible. The aim is to successfully defend the targets even with a small number of defenders, as illustrated in Fig. \ref{fig:bottleneck}.

We attempt to identify strategic bottlenecks and block them by defenders. In order to discover bottlenecks of general shape, we develop the following simulation strategy exploiting the underlying grid graph. The basic idea is that as attackers move towards the targets, they are expected to pass through vertices close to a bottleneck more often than other vertices. This observation suggests to simulate the movement of the attackers and find frequently visited vertices. As defenders do not share the knowledge about paths being followed by attackers, frequently visited vertices are determined by a simulation in which paths of attackers are estimated. 

After obtaining such a frequently visited vertex, we then explore its vicinity up to a given distance. If we find out that there is indeed a bottleneck, its vertices are assigned to some defenders as their new targets. Under the assumption that the bottleneck is blocked by defenders, the paths of attackers may substantially change. For that reason we estimate the paths again and find the next frequent vertex of which vicinity is searched for bottlenecks. The whole process is repeated until all available defenders are allocated to a target, or until no more bottlenecks are found. Alg. \ref{alg:bottleneck} describes this procedure more formally.
\begin{algorithm}
 \KwData{$G=(V,E)$, $D$, $A$}
 \KwResult{Target allocation $\delta^D$}
 $D_{\text{available}} = D$\tcp*{Defenders to be allocated to targets}
 $F = \emptyset$ \tcp*{Set of forbidden locations} 

 $\delta'_A=$ Random guess of $\delta_A$\;
 \While {$D_{\text{available}}\neq \emptyset$}{
  \For{$a\in A$} {
   \tcc{find the shortest path in $G$ between an attacker $a$ and its estimated target, that avoids passing through the forbidden locations in $F$}
  $p_a= \text{shortestPath}(\alpha_0(a), \delta'_A(a), G, F)$\;
}
  $f(v) = |\{p_a:a\in A \wedge v\in p_a\}|$\tcp*{Frequency of $v$}
  $w\in \arg\max_{v\in V} f(v)$\;
  $B=$ exploreVicinity$(w)$\tcp*{Search for a bottleneck}
  \eIf{$B\neq \emptyset$} {
  	$D'\subseteq D_{\text{available}}, |D'|=|B|$\;
    \text{assignToDefenders}(B, D')\;
    $D_{available}=D_{\text{available}}\setminus D'$\;
  	$F = F\cup B$
  }
  {
   \tcc{If no new bottleneck is found, assume all have been already discovered }
   \textbf{break} \;
  }
 }
 \tcc{If there are some defenders without a target left, they will be allocated randomly}
 $\text{assignToRandomTargets}(D_{available})$\;
 ~\newline
 \caption{Bottleneck simulation proceudre}
\label{alg:bottleneck}
\end{algorithm}

\section{\uppercase{Connectivity Maintenance}}

The requirement of maintaining the possibility of communication is modeled by a connectivity maintenance of subgraph of the \emph{visibility graph} induced by the defenders' locations. The first task is therefore to create the visibility graph, which depends on the positions of obstacles in the map and a predetermined visibility range. The agents move using an adaptation of the LRA* algorithm that preserves the connectivity of the visibility graph. Paths are planned such that the first step of a defender must lead to a position which induces a connected visibility graph. Defenders follow paths computed by LRA* and whenever an agent is about to enter an occupied location, or if the next move would disconnect the communication subgraph, its path is recalculated.

The movement determined by such an approach will surely maintain the connectivity, however, in many instances, some defenders will not be able to reach the target locations assigned to them. Our effort is to modify the allocation strategies so that the number of defenders that are not able to reach their assigned targets is minimized.

An intuitive idea assumes that defenders should be allocated to their targets so that in the most optimistic case, when they all reach their targets, the connectivity is preserved. This constraint is not guaranteed to be satisfied in a general case. We propose the following approach to tackle this issue.

\begin{figure}[!h]
  \vspace{-0.2cm}
  \centering
   {\epsfig{file = 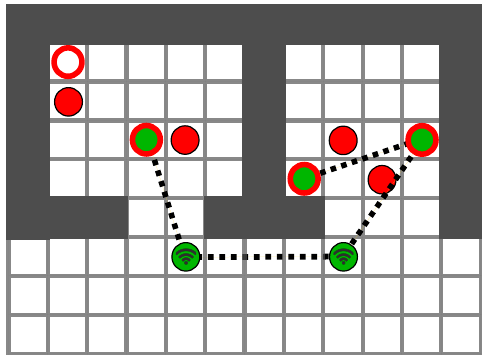, width = 4.0cm}}
  \caption{Three occupiers managed to reach the targets, but due to the wall they are not able to communicate. The presence of two communicators enables the communication via links marked by the dashed lines.  }
  \label{fig:comm-example}
 \end{figure}

Initially, the defenders are partitioned into two sub-sets, \emph{communicators} $D_c$ and \emph{occupiers} $D_o$, with a selected ratio $|D_c|:|D_o|$. The occupiers are allocated to targets according to one of the allocation strategies described in the previous subsections. In the best scenario, all defenders manage to reach their targets. It is easy to check, for example by using BFS or DFS on the induced subgraph, whether the ideal final position of defenders maintains connectivity. If the connectivity is violated, the defenders reserved as communicators are allocated to targets so that the subgraph of the visibility graph induced by the defenders' target locations has as few connected components as possible. Fig. \ref{fig:comm-example} depicts a situation where the three occupiers reached the attackers' targets assigned to them, but they alone are not able to communicate. Nevertheless, a suitable placement of two communicators ensures that the communication can take place.
In fact, the question whether it is possible to allocate targets for $D_c$ so that the desired position allows a communication among all defenders is already difficult.

\begin{proposition}
Let $\Sigma$ be an APPC instance with the set of defenders $D=D_c\cup D_o$. The decision problem whether there exists a target allocation $\delta^{D_2}:D_2\rightarrow V$ of targets to defenders such that all defenders maintain connectivity at their final positions is NP-complete. 
\end{proposition}
\noindent{\emph{Sketch of proof.}}
The problem is obviously NP, because checking a connectivity can be done in polynomial time. In order to prove the NP-hardness, we reduce the NP-complete problem of \emph{Vertex Cover (VC)} to our problem. Let $H=(V_H,E_H)$ be an instance of VC. For each $e\in E_H$ we create $v_e\in V$ such that $V_e$ is a target of some $a\in A$. For each $u\in V_H$ we construct $v_u\in V$ such that for all $e\in E_H$ incident with $u$ we create $\{v_u,v_e\}\in E$. Vertices $v_u$ s. t. $u\in V_H$ form a complete subgraph of $G$.  Finally, set $|D_c|=k$. Now $H$ has a vertex cover of size at most $k$ if and only if it is possible to maintain connectivity of the desired position in $\Sigma$.\qed

Let $T_o$ and $T_c$ be the set of targets allocated to occupiers and communicators, respectively. If the induced subgraph $G_r\left[T_o\right]$ has several  connected components, the used modification of LRA* algorithm could not lead all of them to their targets, because it would cause a loss of communication ability. At this point the set of communicators comes into play. The aim is to find target locations for communicators so that the graph $G_r\left[T_o\cup T_r\right]$ is connected. First, the connected components of $G_r\left[T_o\right]$. are identified. We then iterate while there are available communicators and connected components to be covered by them. In every iteration, a location $l$ from which a communicator can cover a set of connected components that contains maximum number of targets allocated to occupiers is selected together with the set of covered connected components. The location $l$ is subsequently assigned to the closest unallocated communicator. For a more formal explanation see Alg. \ref{alg:communicators}.

\begin{algorithm}
 \KwData{$G_r=(V,E_r)$, $D_o$, $D_c$, $T_o$ }
 \KwResult{Target allocation $\delta^{D_c}$}
$T_c=\emptyset$\tcp*{Targets assigned to communicators}
\While {$D_c\neq \emptyset$} {
$\mathscr{C}$ = connected components of $G_r\left[T_o\cup T_c\right]$\;
\While {$\mathscr{C}\neq \emptyset$} {
\tcc{A pair of a locatoin $l$ and a subset $\mathscr{C}'$ of connected components covered by $l$ that minimizes the number of vertices in $\mathscr{C}'$} 
$(l,\mathscr{C}')=\arg\max\limits_{\mathscr{C}'\in\mathscr{C}, l\in V}\{\sum\limits_{C\in\mathscr{C}'}|C|:\exists v\in C: (v,l)\in E_r\}$\;
\tcc{An available agent closest to $l$}
$a=\arg\min_{a\in D_c}\{|p_{\alpha_0(a),l}|\}$\;
assign the target $l$  to the agent $a$\;
$T_c=T_c\cup \{l\}$\;
$\mathscr{C}=\mathscr{C}\setminus\mathscr{C}'$\;
$D_c=D_c\setminus \{a\}$\;
  \If{$D_c=\emptyset$ }{
  \textbf{break} \;
   }	
}
}
 \caption{Target allocation to communicators}
\label{alg:communicators}
\end{algorithm}

\section{\uppercase{Preliminary experiments }}

The aim of experimental evaluation is to compare individual strategies described in the previous section with their counterparts adapted to connectivity maintenance. We would like to find out whether the adaptation improves the success rate of a strategy and also how instance types affect its performance.

Our hypothesis is that when there is a  sufficient number of defenders, the adaptation has little or no effect. We predict that in instances, where defenders are outnumbered by attackers, the adaptation increases the success rate of the corresponding strategy. Furthermore, it is likely that the simulation strategy is worse when the connectivity maintenance is required, because the identified bottlenecks may be far from each other, which makes it difficult to preserve communication among them.

We implemented all suggested strategies in Java as an experimental prototype. In our testing scenarios we use maps of different structure with various initial configurations of attackers and defenders. Our choice of testing scenarios is focused on comparing performance of the strategies and discovering what factors have impact on their success.

As the following sections show, different strategies are successful in different types of instances. It is therefore important to design the instances with a sufficient diversity, in order to capture strengths and weaknesses of individual strategies. 

\subsection{Instance generation and types}

\begin{figure}[!h]
    \centering
    \begin{subfigure}[b]{0.17\textwidth}
    \centering
        \includegraphics[height = 1.7cm]{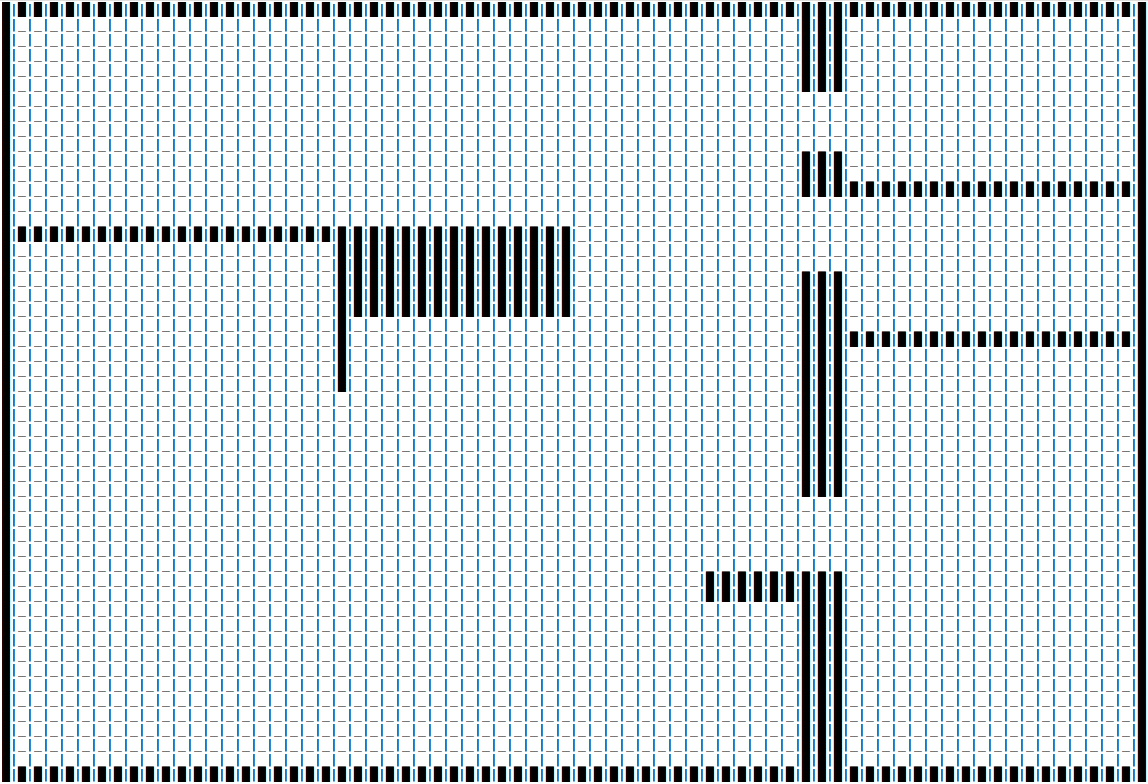}
        \caption{Orthogonal rooms}
        \label{fig:orthogonal_rooms}
    \end{subfigure}
    \begin{subfigure}[b]{0.15\textwidth}
    \centering
        \includegraphics[width = 1.7cm, angle = 90]{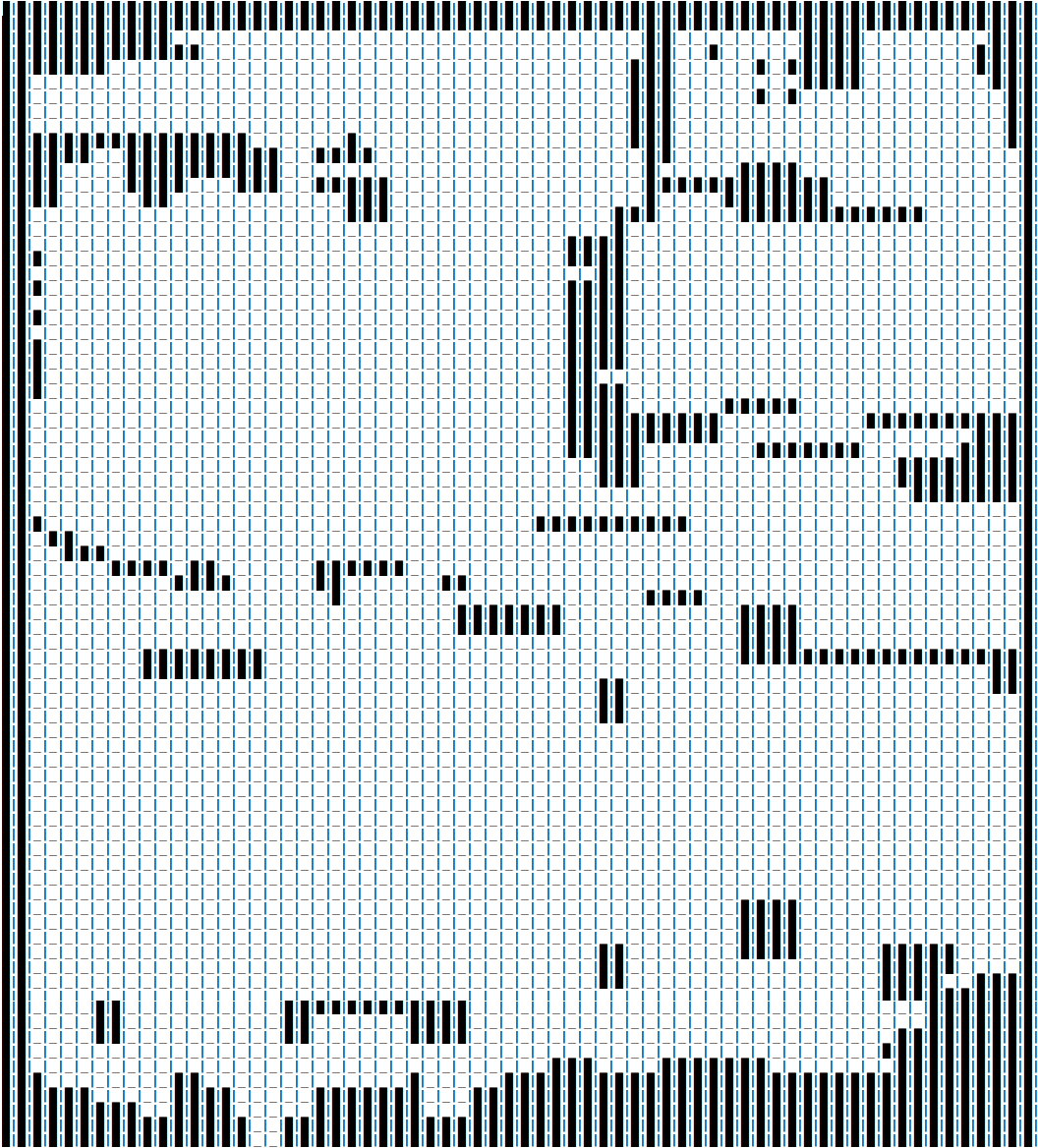}
        \caption{Ruins}
        \label{fig:ruins}
    \end{subfigure}
    \vspace{3mm}
    \begin{subfigure}[b]{0.15\textwidth}
    \centering
        \includegraphics[height = 1.7cm]{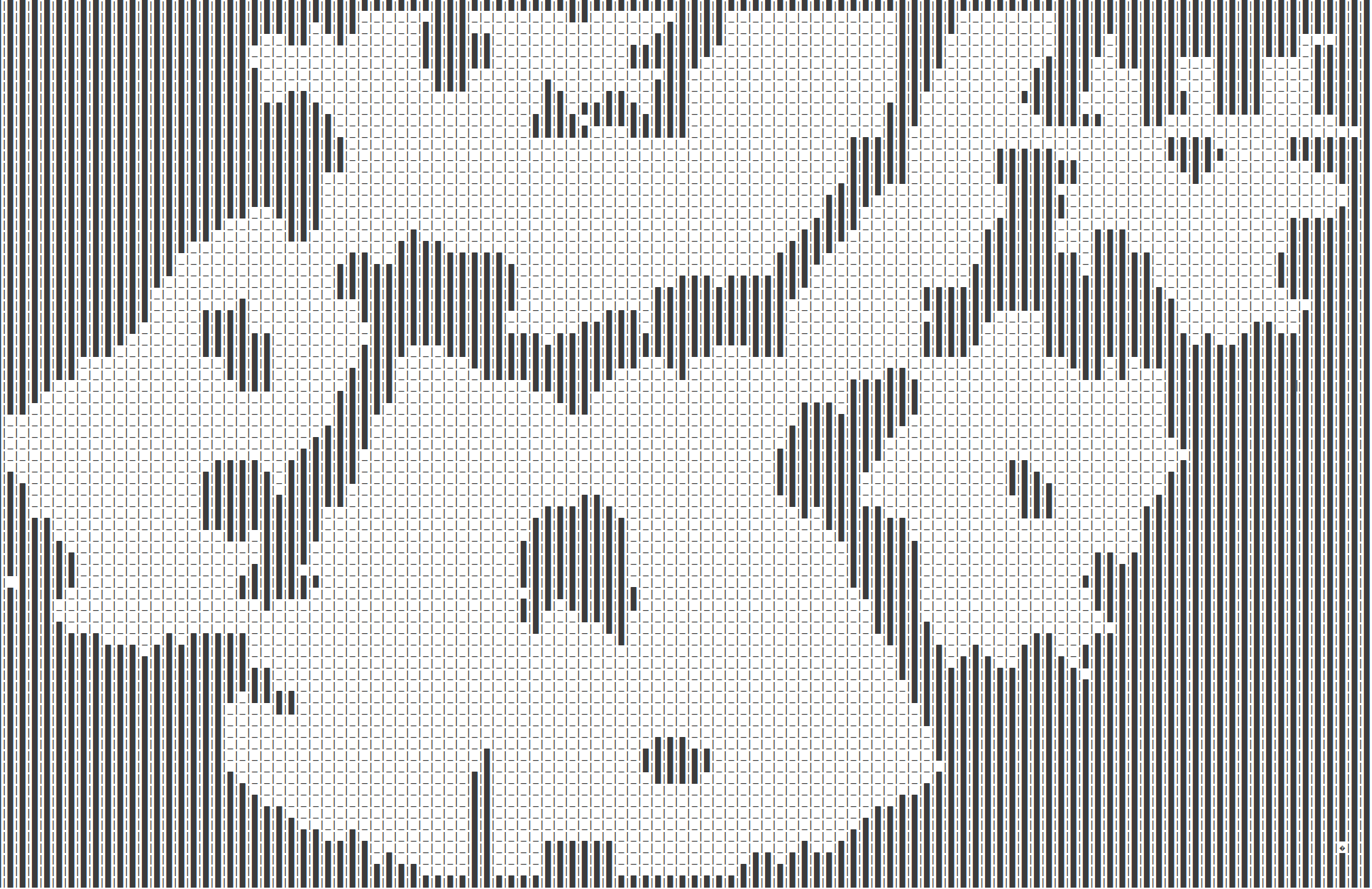}
        \caption{Waterfront}
        \label{fig:waterfront}
    \end{subfigure}
        \vspace{3mm}
    \caption{Three different maps used in the evaluation}\label{fig:maps}
\end{figure}

The instances used in the practical experiments are generated using a pseudo random generator, but in a controlled manner. An instance is defined by its map, the ratio $|A|:|D|$ and locations of individual defenders, attackers and their targets. These three entries form an input of the instance generation procedure. Further, we select rectangular areas inside which agents of both teams and the attackers' targets are placed randomly. The experiments are conducted on 3 different maps that vary in their structure. The maps are depicted in Fig. \ref{fig:maps}. 




 

Each map is populated with agents of 3 different $|D|:|A|$ ratios, namely $1:1$, $1:2$ and and $1:5$, with fixed number of attackers $|A|=50$. The maximum number of moves of the agents is set to 150 for each team.
Note that the individual instances are never completely fair to both teams. It is therefore impossible to make a conclusion about a success rate of a strategy by comparing its performance on different maps. The comparison should always be made by inspecting the performance in one type of instance, where we can see the relative strength of the studied algorithms.

\subsection{Experimental results}
The following set of experiments compares random, greedy, simulation strategy and their communication counterparts in different instance settings. Each of the following tables contains results for one map.

Each entry in following tables shows an average number of attackers that reached their targets at the end of the time limit. The average value is calculated for 10 runs in each settings, always with a different random seed. Random and greedy strategies have very similar results in all positions and team ratios. It is apparent and not surprising that with decreasing $|D|:|A|$ ratio, the strength of defensive strategies decreases. 

 \begin{table}[h]
	\caption{Average number of agents that eventually reached their target in the map Orthogonal rooms}\label{tab:orthogonal-rooms} \centering
\begin{tabular}{crrrrrr}
	 & \multicolumn{6}{c}{Strategies}\Bstrut\\ \hline
     $|D|:|A|$ & RND & RND-C & GRD&  GRD-C & SIM & SIM-C \TBstrut\\ \hline
     \multicolumn{1}{c|}{1:1} & 26.0 & \multicolumn{1}{r|}{29.0} & 25.5&  \multicolumn{1}{r|}{29.1} & 20.8 & 28.3 \TBstrut\\
     \multicolumn{1}{c|}{1:2} & 41.0 & \multicolumn{1}{r|}{39.6} & 39.4&  \multicolumn{1}{r|}{40.5} & 29.3 & 31.7 \TBstrut\\
     \multicolumn{1}{c|}{1:5} & 48.1 & \multicolumn{1}{r|}{45.7} & 46.1&  \multicolumn{1}{r|}{46.8} & 46.9 & 46.8 \TBstrut\\
\end{tabular}
\end{table}
We focused on evaluation of the effect of using communicating agents in implemented target allocation strategies. For each target allocation strategy we compare the standard version and the version with communicating agents.

Tab. \ref{tab:orthogonal-rooms} shows results for Orthogonal rooms map. It can be observed that using communicators is beneficial in case of random strategy where defenders tend to be outnumbered by attackers. On the other hand, communicators cause no improvement in Ruins map (Tab. \ref{tab:ruins}).
\begin{table}[h]
	\caption{Average number of agents that eventually reached their target in the map Ruins. }\label{tab:ruins} \centering
\begin{tabular}{crrrrrr}
	 & \multicolumn{6}{c}{Strategies}\Bstrut\\ \hline
     $|D|:|A|$ & RND & RND-C & GRD&  GRD-C & SIM & SIM-C \TBstrut\\ \hline
     \multicolumn{1}{c|}{1:1} & 21.5 & \multicolumn{1}{r|}{21.1} & 24.8&  \multicolumn{1}{r|}{24.7} & 18.3 & 18.6 \TBstrut\\
     \multicolumn{1}{c|}{1:2} & 42.1 & \multicolumn{1}{r|}{40.2} & 39.0&  \multicolumn{1}{r|}{40.3} & 37.1 & 36.9 \TBstrut\\
     \multicolumn{1}{c|}{1:5} & 47.1 & \multicolumn{1}{r|}{47.1} & 46.0&  \multicolumn{1}{r|}{46.2} & 44.3 & 43.8 \TBstrut\\
\end{tabular}
\end{table}
Small improvement of the bottleneck simulation strategy can be observed in Waterfront map (Tab. \ref{tab:waterfront}) again in cases when defenders are outnumbered. Both types of maps where communicators turned out to be beneficial appear to have the structure of large open spaces separated by narrow bottlenecks.
 \begin{table}[h]
	\caption{Average number of agents that eventually reached their target in the map Waterfront}\label{tab:waterfront} \centering
\begin{tabular}{crrrrrr}
	 & \multicolumn{6}{c}{Strategies}\Bstrut\\ \hline
     $|D|:|A|$ & RND & RND-C & GRD&  GRD-C & SIM & SIM-C \TBstrut\\ \hline
     \multicolumn{1}{c|}{1:1} & 20.7 & \multicolumn{1}{r|}{21.6} & 18.9&  \multicolumn{1}{r|}{18.5} & 20.8 & 21.9 \TBstrut\\
     \multicolumn{1}{c|}{1:2} & 35.2 & \multicolumn{1}{r|}{31.2} & 30.7&  \multicolumn{1}{r|}{31.4} & 35.8 & 33.5 \TBstrut\\
     \multicolumn{1}{c|}{1:5} & 41.6 & \multicolumn{1}{r|}{41.4} & 40.7&  \multicolumn{1}{r|}{40.7} & 42.3 & 41.3 \TBstrut\\
\end{tabular}
\end{table}
\section*{\uppercase{Conclusion and future work}}
We have designed several practical algorithms for APPC. We extended previous algorithms for APP with a technique of connectivity maintenance. This is done by dividing defending agents into two groups - occupiers and communicators. The role of occupiers is to protect the area while communicators are placed so that they cover as largest part of the protected area as possible in order to support connectivity among occupiers. 
Performed experimental evaluation indicates that the effect of using dedicated agents as communicators is much smaller than expected but there is some in maps having the structure of large open spaces separated by bottlenecks. One possible explanation of this behavior is that several defenders are not able to reach their targets because the ability of  communication would be lost during their movement and this is not significantly affected by the target allocation.  Hence, for the future work we plan to design and evaluate algorithms with more sophisticated mechanism for connectivity maintenance. A more promising direction seems to be an adaptation of LRA* rather than modifications of the allocation strategies.


\bibliographystyle{IEEEtran}
\bibliography{connected-agents}

\vfill
\end{document}